# High Field Magneto-Conductivity Analysis of $Bi_2Se_3$ Single Crystal

Rabia Sultana[1,2], Ganesh Gurjar[3], S. Patnaik[3] and V.P.S. Awana[1,2*]

[1] National Physical Laboratory (CSIR), Dr. K. S. Krishnan Road, New Delhi-110012, India

[2] Academy of Scientific and Innovative Research (AcSIR), NPL, New Delhi-110012, India

[3] School of Physical Sciences, Jawaharlal Nehru University, New Delhi-110067, India

**Abstract**

We report the high field (up to 14Tesla) magneto-conductivity analysis of $Bi_2Se_3$ topological insulator grown via the self flux method. The detailed experimental investigations including crystal growth as well as the electrical, thermal and spectroscopic characterizations of the resultant $Bi_2Se_3$ single crystal are already reported by some of us. The current letter deals with high field magneto-conductivity analysis in terms of Hikami Larkin Nagaoka (HLN) model, which revealed that the electronic conduction is dominated by both surface states driven weak anti localization (WAL), as well the bulk WL (weak localization) states. Further, by applying the HLN equation we have extracted the fitting parameters i.e., phase coherence length ($l_\varphi$) and the pre-factor (α). Here, the magneto-conductivity data is fitted up to ± 5Tesla, but in order to extract reliable fitting parameters the same is fitted at much lower magnetic fields i.e., up to ± 1Tesla. The value of the HLN coefficient (α), extracted from the HLN equation exhibited values close to -1.0, indicating both WAL and WL contributions. On the other hand, the extracted phase coherence length ($l_\varphi$) is seen to decrease from 11.125nm to 5.576nm as the temperature is increased from 5K to 200K respectively. Summarily, the short letter discusses primarily the temperature dependent magneto-conductivity analysis of pristine $Bi_2Se_3$ single crystal by the HLN model.

*Corresponding Author
Dr. V. P. S. Awana: E-mail: awana@nplindia.org
Ph. +91-11-45609357, Fax-+91-11-45609310
Homepage: awanavps.webs.com

**Introduction**

Three dimensional (3D) layered materials popularly known as the topological insulators (TIs) are one of the novel materials discovered recently [1-3]. $Bi_2Te_3$, $Bi_2Se_3$ and $Sb_2Te_3$ are among the most extensively studied 3D TI materials due to their single surface Dirac cone along with a relatively large bulk energy gap [1- 4]. These layered materials represent a new phase of quantum matter exhibiting simultaneously bulk band gap in the interior and the gapless electronic surface states protected by time reversal symmetry (TRS) [1-6]. Historically, these TIs were known to be good thermoelectric materials but owing to their exotic surface states and unusual magnetic properties, these now hold great promises for novel applications such as spintronics and quantum computers owing to fascinating quantum effects and topological order [1-6]. The existence of the topological surface states has been experimentally confirmed by the angle-resolved photoemission spectroscopy (ARPES) and scanning tunneling microscopy (STM) measurements [4-6]. Indeed, various quantum transport phenomenon viz. weak anti localization (WAL), quantum conductance fluctuation, Aharonov-Bohm (AB) oscillations and high field linear magneto-resistance (MR) associated with the topological surface states are already attracting huge attention of scientific community [7-14]. Among these quantum transport phenomena of TIs, the WAL effect arising from the π- Berry phase of the topological surface states act as one of the main driving force behind the unusual MR as well various other intriguing transport feature of TIs [9-14].

However, the electronic transport measurements still remains a challenge, because the surface states driven WAL is often dominated by the WL bulk contribution to the conduction process. This precisely occurs due to the opening of a surface state gap from say intrinsic defects or unintentional doping giving rise to weak localization (WL) effect [9-13]. Both WAL and WL are the quantum interference effects, which respectively lead to the enhancement and suppression of the conductivity at lower temperatures. However, in TI system the quantum interference effect can be easily destroyed by the application of small magnetic field. WAL appears with the negative magneto-conductivity with a sharp dip (cusp) in lower applied fields, which is well in a system with strong spin orbit coupling (SOC). On the other hand WL is seen as cusp like positive magneto-conductivity at lower applied field and temperatures [9-13].

Experimentally, the analysis of WAL effect in 3D TIs is done by fitting the magneto-conductivity data using the well known Hikami Larkin Nagaoka (HLN) equation [15]. In the present work, we discuss and analyze our reported high field linear non saturating MR behavior of pristine $Bi_2Se_3$ single crystal [16], in terms of the HLN equation and do conclude that in TIs often both conducting surface driven WAL and insulating bulk dominated WL do contribute to the overall conductivity, a result very similar to that as reported by us recently for $Bi_2Te_3$ [13]. This is further consistent with our recent observations related to unexplored photoluminescence of $Bi_2Te_3$, showing that TIs do exhibit electronic transitions from bulk insulating interior to conducting surface states [17].

**Experimental details**

Conventional self flux method was used to synthesize quality bulk single crystals of $Bi_2Se_3$. The experimental technique including the crystal growth and the heat treatment diagram along with the structural, micro-structural, spectroscopic, and thermal properties are described in detail in our own past work [16]. As the TIs exhibit a layered structure, the mechanical cleaving was easily performed and the easily cleaved thin flakes of $Bi_2Se_3$ single crystal revealed a silvery shining mirror like flat surface. The magneto transport measurements were carried out on one the cleaved flake using 14Tesla down to 2K, Quantum Design Physical Property Measurement System (PPMS), Model 6000 equipped with standard four probe technique. Silver paste was employed for making electrical transport contacts on freshly cleaved shining surface of the crystal.

**Results and Discussion**

Figure 1 displays the single crystal XRD (X-ray diffraction) pattern of as synthesized $Bi_2Se_3$ recorded using the Rigaku-Miniflex II having Cu- $K_\alpha$ radiation ($\lambda$=1.5418 Å). The X-ray analysis was carried out on the cleaved silvery surface of the $Bi_2Se_3$ single crystal. The XRD diffraction spectra revealed highly indexed peaks with (00l) reflections indicating the crystalline nature of the synthesized $Bi_2Se_3$ crystal. Also to be noted that tiny peaks around 18.5 and 30 degree do appear, the later one is due to misaligned [015] plane of $Bi_2Se_3$, and former is not identified. The said peaks are marked in the Fig 1 as well. Further, the powder XRD patterns confirmed the rhombohedral crystal structure of the as synthesized $Bi_2Se_3$ single crystal; more details are given in ref. 16.

The inset of Fig.1 shows the typical percentage change of MR of as grown $Bi_2Se_3$ single crystal as a function of perpendicular applied magnetic fields at different temperatures. Here, the MR is calculated using the equation MR (%) = {[$\rho$(H) - $\rho$(0)] / $\rho$(0)}*100, where  $\rho$(0) and ρ(H) represents the resistivity values under zero and non zero applied magnetic fields (H) respectively. Apparently, it can be seen that all the MR curves exhibit positive MR values. A linear, non-saturating MR behavior reaching up to a value of about 380% is observed under 14Tesla and is maintained nearly the same at temperature below 20K. At 50K, 100K and 200K, the MR reaches a value of about 280%, 150% and 60% respectively under 14Tesla applied magnetic field. Also, the shape of the MR curve tends to broaden as the temperature is increased from 5K to 200K. The quantitative MR (%) values at different temperatures under 14Tesla applied magnetic of as grown $Bi_2Se_3$ single crystal are listed in Table 1.

Now to understand the MR behavior, shown in inset of Fig.1, we employed the HLN equation, which is the main output of the present letter. As discussed, the WAL effect arises from the topological surface states and appears as a negative magneto-conductivity cusp (sharp dip) at lower fields. However, with increase in temperature and magnetic field, the sharp cusp type behavior gradually gets suppressed. This suppression of WAL behavior occurs due to the breaking of TRS induced by magnetic field. In order to analyze the quantum corrections to the conductivity of pristine $Bi_2Se_3$ single crystal experimentally, the magneto-conductivity curves are fitted using the standard Hikami Larkin Nagaoka (HLN) equation; [15]

$$\Delta\sigma(H) = \sigma(H) - \sigma(0) = -\frac{\alpha e^2}{\pi h}\left[\ln(\frac{B_\varphi}{H}) - \Psi\left(\frac{1}{2} + \frac{B_\varphi}{H}\right)\right]$$

Where, $\Delta\sigma(H)$ represents change of magneto-conductivity, $\Psi$ is the digamma function, e is the electronic charge, h is the Planck's constant, $B_\varphi = \frac{h}{8e\pi H l_\varphi}$ is the characteristic magnetic field, H is the applied magnetic field. Here, $l_\varphi$ and α are the two fitting parameters signifying phase coherence length and pre factor respectively. The pre factor, α is a coefficient indicating the type of localization (WL, WAL or both WL and WAL) and exhibits values depending upon the type of spin orbit interaction (SOI) and magnetic scattering [15]. Accordingly, α = 0 when the magnetic scattering is strong, α = 1 when the SOI and magnetic scattering is weak or absent and α = -1/2 when SOI is strong and there is no magnetic scattering [15].

Figure 2 depicts the standard HLN fitting of magneto-conductivity curves for $Bi_2Se_3$ single crystal carried out at various temperatures under an applied magnetic field of ± 5Tesla. As expected, a negative change in conductance is observed (figure 2 and 3). The symbols in figure 2 and 3 represent the experimentally observed data, whereas solid lines correspond to the HLN fitting curves. At lower temperatures (5K, 10K and 20K) the magneto-conductivity curves are observed to overlap on each other whereas, distinct magneto-conductivity curves are seen at temperatures above 20K. Figure 2 clearly shows that as the temperature increases from 5K to 200K the magneto-conductivity curves tend to exhibit less negative values. In consequence, all the magneto-conductivity curves follow HLN behavior.

In order to extract reliable fitting parameters ($\alpha$ and $l_\varphi$) and to study the HLN equation more specifically the magneto-conductivity curves at different temperatures are HLN fitted at much lower magnetic fields i.e., up to $\pm$ 1Tesla as depicted in figure 3. The magneto-conductivity curves fits the standard HLN equation quite well and yields reasonable values of the fitting parameters ($\alpha$ and $l_\varphi$). The fitting parameters ($\alpha$ and $l_\varphi$) are provided in a tabular form (Table 1) as well as in the figure 3 itself. Ideally, the HLN coefficient ($\alpha$), should be -1/2 for each transport channel that carries a $\pi$ Berry phase or exhibits strong SOI whereas, for multiple carrier channels (surface and bulk states) the value should add up to bigger negative value viz. -1 and -1.5. Usually, the experimentally fitted value of $\alpha$ covers a wide range from -0.4 to -1.1, suggesting that the observed WAL effect arises from single surface state, two surface states or intermixing between the WAL of the surface states and WL of the bulk states. Accordingly, it becomes difficult to differentiate the conduction taking place from bulk and surface states clearly. Additionally, the fitting parameter ($\alpha$ and $l_\varphi$) values depends from material to material viz., bulk single crystals, thin films, nano wires and thin flakes [9-13].

In our case, the pre-factor, $\alpha$ measuring the strength of WAL is observed to exhibit values close to around -1, suggesting that both surface states (WAL) and bulk states (WL) contribute to the conduction mechanism. Accordingly, the extracted value of $\alpha$ is close to the predicted theoretical value of -1 for all temperatures (5K, 10K, 20K, 50K, 100K and 200K). Indeed, one can see from table 1, that the obtained value of $\alpha$ remains nearly unchanged as a function of temperature. However, phase coherence length ($l_\varphi$) is observed to decrease from 11.125 nm to 5.576 nm at 5K and 200K respectively. This decrease in phase coherence length as a function of temperature is expected since the probability of inelastic scattering due to electron-electron interaction increase with temperature. Recently, in case of $Bi_2Te_3$ bulk single crystal as well we found good fitting of magneto-conductivity data to HLN equation with pre factor ($\alpha$) close to -0.9 and phase coherence length ($l_\varphi$) decreasing from 98nm to 40nm with increase in temperature from 2.5K to 50K [13]. Further, this is consistent with our recent results related to unexplored photoluminescence from bulk and mechanically exfoliated few layers of $Bi_2Te_3$, exhibiting electronic transition from bulk insulating to surface conducting states [17]. The HLN analysis of high field magneto conductivity data of both $Bi_2Te_3$ [13] and presently studied $Bi_2Se_3$ show the contributions from both WL (bulk insulating) and WAL (surface conducting) states to overall conduction of these composite systems.

Summarily, it is seen that the magneto conductivity of the single crystalline $Bi_2Se_3$ topological insulator follows the standard HLN model being dominated by both surface and bulk states driven WAL and WL processes.

**Acknowledgements**

The authors from CSIR-NPL would like to thank their Director NPL, India, for his keen interest in the present work. S. Patnaik thanks DST-SERB project (EMR/2016/003998) for the low temperature high magnetic facility at JNU, New Delhi. Rabia Sultana and Ganesh Gurjar thank CSIR, India, for research fellowship. Rabia Sultana thanks AcSIR-NPL for Ph.D. registration.

**Figure Captions**

**Figure 1:** X-ray diffraction pattern of as synthesized $Bi_2Se_3$ single crystal. Inset shows the percentage change of magneto-resistance at different temperatures for $Bi_2Se_3$ single crystal.

**Figure 2:** WAL related magneto-conductivity for $Bi_2Se_3$ single crystal at different temperatures (5K, 10K, 20K, 50K, 100K and 200K), fitted using the HLN equation up to ± 5Tesla.

**Figure 3:** Magneto-conductivity curves for $Bi_2Se_3$ single crystal at different temperatures (5K, 10K, 20K, 50K, 100K and 200K), fitted using the HLN equation up to ± 1Tesla

**Table 1** MR (%) values at different temperatures under 14Tesla applied magnetic field as well as HLN fit values of pre factor (α) and phase coherence length ($l_\varphi$) for $Bi_2Se_3$ crystal.

Table 1

| Temperature (K) | α | $l_\varphi$ (nm) | MR (%) at 14Tesla |
|---|---|---|---|
| 5 | -0.9897 | 11.125 | ∼ 380 |
| 10 | -1.040 | 10.895 | ∼ 380 |
| 20 | -1.045 | 10.667 | ∼ 380 |
| 50 | -1.012 | 9.496 | ∼ 280 |
| 100 | -0.999 | 7.805 | ∼ 150 |
| 200 | -0.996 | 5.576 | ∼ 60 |

**Fig.1**

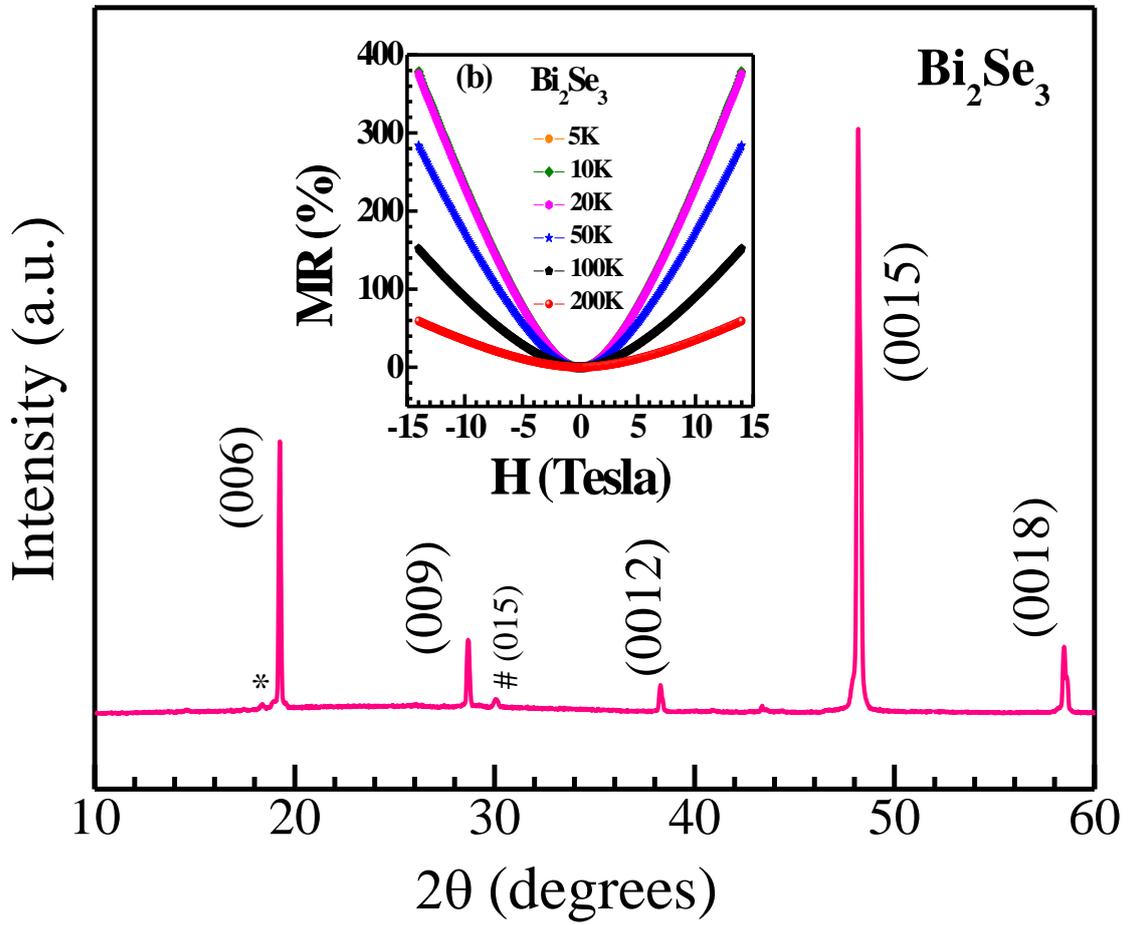

**Fig.2**

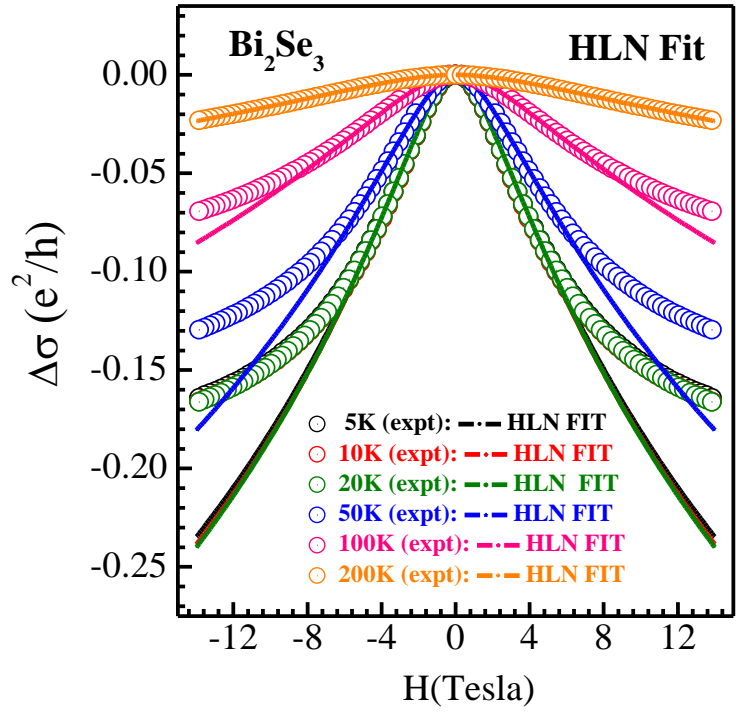

**Fig. 3**

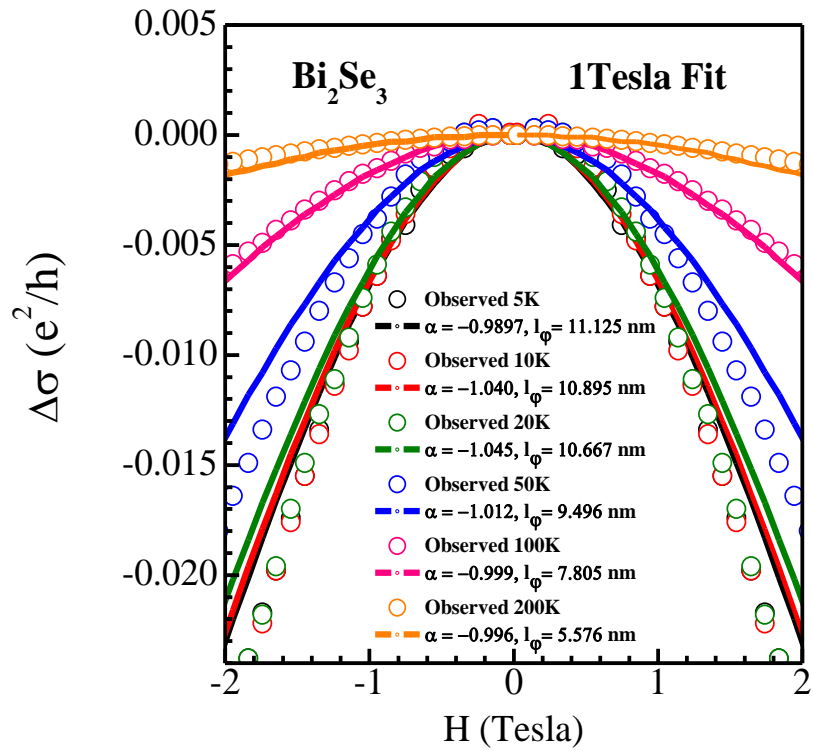